\documentclass[11pt,twoside]{article}
\usepackage{./asp2014}
\usepackage{xcolor}

\aspSuppressVolSlug
\resetcounters

\begin{document}

\title{Disk Winds and the Evolution of Planet-Forming Disks}
\author{I. Pascucci,$^1$ S. Andrews,$^2$ C. Chandler,$^3$ and A. Isella$^4$
\affil{$^1$Lunar and Planetary Laboratory, Tucson, AZ, USA; \email{pascucci@lpl.arizona.edu}}
\affil{$^2$Harvard-Smithsonian Center for Astrophysics, Cambridge, MA, USA; \email{sandrews@cfa.harvard.edu}}
\affil{$^3$National Radio Astronomy Observatory, Socorro, NM, USA; \email{cchandle@nrao.edu}}
\affil{$^4$Rice University, Houston, TX, USA; \email{isella@rice.edu}}}

\paperauthor{Sample~Author1}{Author1Email@email.edu}{ORCID_Or_Blank}{Author1 Institution}{Author1 Department}{City}{State/Province}{Postal Code}{Country}
\paperauthor{Sample~Author2}{Author2Email@email.edu}{ORCID_Or_Blank}{Author2 Institution}{Author2 Department}{City}{State/Province}{Postal Code}{Country}
\paperauthor{Sample~Author3}{Author3Email@email.edu}{ORCID_Or_Blank}{Author3 Institution}{Author3 Department}{City}{State/Province}{Postal Code}{Country}

\begin{abstract}
Disk winds are thought to play an important role in the evolution and dispersal of planet-forming disks. While high-resolution optical and infrared spectroscopy has identified several disk wind diagnostics, wind mass loss rates remain largely unconstrained mostly due to the lack of spatial resolution to measure the extent of the wind emitting region. Here, we show that the ngVLA will have the sensitivity and resolution to detect and spatially resolve the free-free emission from the fully or partially ionized component of  disk winds. Hence, ngVLA observations will be critical to estimate mass loss rates and clarify the role of disk winds in the evolution and dispersal of disk mass.
\end{abstract}

\section{Introduction}
Disks of gas and dust around young ($\sim 1-10$\,Myr) stars are active sites of planet formation. Hence, their  evolution and dispersal directly impact which type of planetary systems can form. Accretion through the circumstellar disk is a ubiquitous phenomenon (see \citealt{hartmann2016} for a recent review) and thought to dominate disk evolution at early times (e.g., \citealt{alexander2014}). However,  the physical mechanism enabling accretion, and therefore global disk evolution, is as yet unclear  (e.g., \citealt{turner2014}). 

The prevailing view for the past 40 years has been that accretion  occurs because disks are viscous and transport angular momentum outward via correlated turbulent fluctuations. Magneto-rotational instability (MIR, \citealt{bh1998}) would produce the necessary turbulence. In this picture, viscous-dominated evolution  lasts for the first few Myr of disk evolution 
until accretion drops below the mass loss rate from thermal disk winds driven by high-energy stellar photons, also known as photoevaporative winds. At that point photoevaporation limits the supply of gas to the inner ($\sim 1$\,au) disk  which drains onto the star on the local viscous timescale, of order 100,000 years (e.g., \citealt{gorti2016}). Recently, simulations including non-ideal MHD effects have shown that most of the disk is not MRI active, hence not accreting onto the star (e.g., \citealt{bs2013}). Instead, these simulations develop vigorous magneto-thermal MHD winds that extract angular momentum from the disk surface and hence enable accretion (e.g., \citealt{gressel2015}). In this scenario, wind mass loss rates are comparable to mass accretion rates even at early times and disk dispersal can be rapid if the disk retains most of its magnetic flux during evolution (\citealt{bai2016} but see also \citealt{zs2017}). As the surface density evolution of a viscous disk is very different from that of a disk where accretion is driven by MHD winds, and surface density directly impacts planet formation, it is important to understand which physical process dominates. Detecting disk winds, spatially resolving them, and measuring wind mass loss rates is crucial to make progress in this field.


\section{Disk wind diagnostics - the role of cm observations}
Identifying disk winds and understanding their origin requires detecting gravitationally unbound/outflowing disk gas within a $\sim$30\,au radius from the star (e.g., \citealt{simon2017}).  Given the typical distance of star-forming regions ($\sim$140\,pc), this requirement translates into a spatial resolution better than $\sim$100\,mas. This is why most direct evidence for disk winds relies on spatially unresolved/high-resolution ($\Delta v \sim$10\,km/s) optical and infrared spectroscopy of gaseous species probing the disk surface (see \citealt{ep2017} for a recent review). Kinematic separation of different components in optical forbidden lines has demonstrated that MHD disk winds are present in the inner $\sim$0.5\,au for the majority of accreting stars while line ratios  constrain the range of temperature-electron densities for the emitting gas (e.g., \citealt{simon2016}). Infrared forbidden line profiles hinting to more radially extended disk winds that could be photoevaporative in nature have been also detected toward a dozen disks (e.g., \citealt{ps2009}, \citealt{sacco2012}).  {\it The missing critical parameter for a reliable computation of wind mass loss rates is a measurement of the spatial extent of the wind emitting region.} 

As a fully or partially ionized disk surface emits free-free continuum radiation and H recombination lines, long mm and cm-wavelength observations can be also used to detect disk winds \citep{pascucci2012,owen2013}.
Current radio facilities have identified candidate disk wind emission from the presence of emission in excess of the thermal dust emission and, when available, multi-wavelength cm observations have been used to exclude other possible sources of excess emission, such as gyro synchrotron non-thermal radiation and emission from very large cm-size grains or very small nanometer-size grains (e.g., \citealt{pascucci2014}, \citealt{galvan-madrid2014}). Even in these multi-wavelength studies, it remains difficult to properly separate free-free emission from a collimated fast ($\sim$100\,km/s) jet (e.g., \citealt{anglada1998}) and from the ionized disk surface, which hampers measuring the disk ionization fraction and the wind mass loss rate. This is illustrated in Figure~\ref{fig:gm} which reports the 3.3\,cm VLA image of GM~Aur \citep{macias2016}, a star surrounded by a disk with a dust cavity (e.g., \citealt{andrews2011,oh2016}) . With a spatial resolution of $0.5''$, $\sim 70$\,au at the distance of Taurus-Auriga, an elongation perpendicular  
to the disk emission is barely detected. This elongation strongly suggests that a jet contributes to the excess cm emission even in disks that are thought to be substantially evolved. {\it Higher spatial resolution and sensitivity are necessary to separate the jet and disk wind contributions.}


\articlefigure{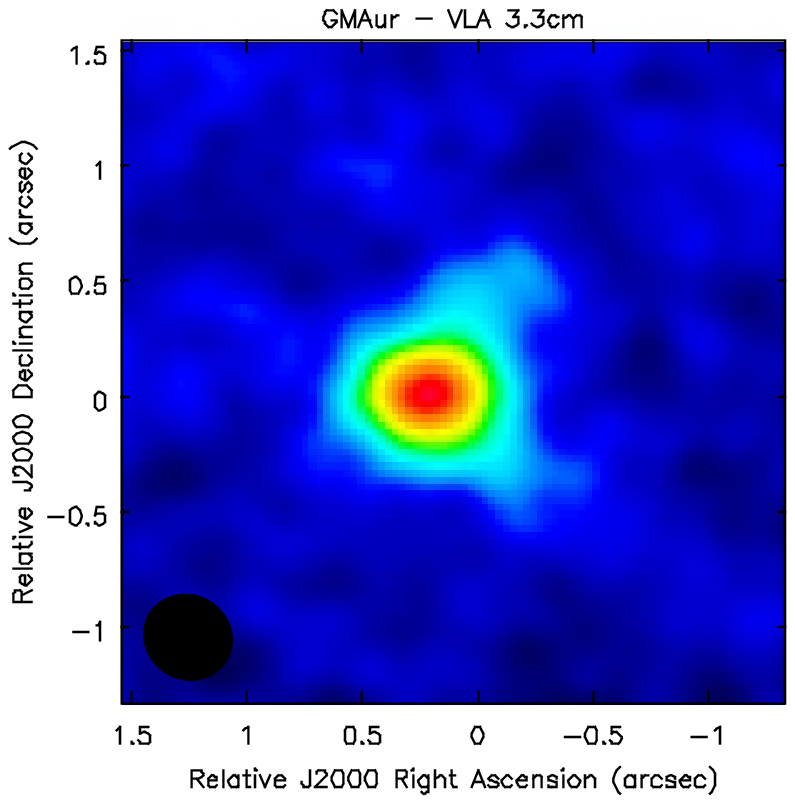}{fig:gm}{ VLA 3.3\,cm image of GM~Aur by \citet{macias2016}. The north-west elongation is attributed to free-free emission from an accretion driven radio jet.}

\section{Connection to the unique ngVLA capabilities}
Spatially resolving disk winds requires deep, multi-frequency radio continuum imaging at high angular resolution, and supplementary spectral imaging of H recombination lines. The continuum measurements target
free-free emission from an ionized wind (and potentially jet). As dust thermal emission dominates
at frequencies higher than 30\,GHz, modest spectral coverage ($\Delta \nu / \nu \sim 0.8$) in the 5-30\,GHz range is essential. For a photoevaporative wind heated and ionized solely by stellar X-rays  and typical star/disk parameters the
expected integrated flux density at 8\,GHz is $\sim$3$\mu$Jy (eq.~3 in \citealt{pascucci2012})\footnote{An EUV ionized layer would produce more free-free emission (eq.~2 in \citealt{pascucci2012}) but observations suggest that the disk receives only a small fraction of the stellar EUV luminosity  \citep{pascucci2014}.} and should arise within a $\sim$10\,au radius (Fig.~4 in \citealt{owen2013}). Spatially resolving the emission in two beams and requiring a S/N of at least 5 means reaching an RMS noise level of 0.3$\mu$Jy/beam. Theoretical estimates for the free-free continuum (and H recombination line emission) from
MHD winds have not yet been reported in the literature. However, observations tracing the jet and MHD wind
 of DG~TauA hint at higher flux densities ($\sim$300$\mu$Jy) distributed over the $\sim 0.1''$ beam of e-MERLIN  (upper panel of Fig.~1 in \citealt{ainsworth2013}).
 

In a fully or partially ionized optically thin region, such as a photoevaporative wind, the H line to free-free continuum ratios increase with frequency but so does the thermal dust
emission. Unless the thermal continuum and H lines can be spatially separated, the best frequency to detect H lines is around 30\,GHz (lower panel of Fig.~2 in \citealt{pascucci2012}). At these
frequencies, the integrated line flux densities from a photoevaporative wind are comparable to the free-free continuum emission ($\sim 3 \mu$Jy) and a few percent of the total continuum (free-free + thermal dust), middle and lower panels of Fig.~2 in \citet{pascucci2012}. Line widths and blueshifts with respect to the stellar velocity depend on disk inclination but typical values are $\sim 10$\,km/s (e.g.,  \citealt{alexander2008,eo2010})\footnote{If H recombination lines would trace an MHD wind launched inside $\sim$1\,au line widths and blueshifts would be substantially larger (e.g., \citealt{romanova2009}).}. Such line emission is probably too weak to detect even with the ngVLA. However, if the line emission from the jet has a similar flux density as the continuum it would detectable, and be particularly important for tracing the accretion and mass-loss
in embedded protostars that are obscured at optical and infrared wavelengths. 
With 300$\mu$Jy distributed over 10\,ngVLA beams obtaining  a 5-sigma detection in a 5\,km/s channel (to provide kinematic information) would take a few 10s of hours with the ngVLA; stacking multiple lines (all available in the 30 GHz
band) would bring the time required down by a factor of the number of lines stacked.


\section{Conclusions}
In this contribution we have illustrated that the sensitivity and spatial resolution of the ngVLA can bring the study of disk winds to the next level. Although H recombination lines will remain challenging to detect, the free-free continuum emission of even a partially ionized disk surface can be separated by that of a jet and, for the first time, imaged. 

The combination of collecting area and sensitivity at 30\,GHz needed for this science is best matched by the ngVLA and will not be available with any other facility in the world. Although ALMA is expected to add Band-1, and
perhaps array receivers, it will have the same sensitivity as the current VLA for point-source
observations and the best resolution will be just over 100\,mas. SKA is not currently expected to operate at frequencies as high as 30 GHz.




\end{document}